# Lessons Learned and Results from Applying Data-Driven Cost Estimation to Industrial Data Sets

J. Heidrich, A. Trendowicz, J. Münch
*Fraunhofer IESE*
*Fraunhofer-Platz 1*
*67663 Kaiserslautern,*
*Germany*
heidrich@iese.fhg.de,
trend@iese.fhg.de,
muench@iese.fhg.de

Y. Ishigai, K. Yokoyama, N. Kikuchi
*IPA-SEC*
*2-28-8 Honkomagome,*
*Bunkyo-Ku, Tokyo, 113-6591,*
*Japan*
ishigai@ipa.go.jp,
k-yokoya@ipa.go.jp,
n-kiku@ipa.go.jp

T. Kawaguchi
*Toshiba Information Systems*
*(Japan) Corporation,*
*7-1 Nissin-Cho,*
*Kawasaki-City 210-8540,*
*Japan*
kawa@tjsys.co.jp

## Abstract

*The increasing availability of cost-relevant data in industry allows companies to apply data-intensive estimation methods. However, available data are often inconsistent, invalid, or incomplete, so that most of the existing data-intensive estimation methods cannot be applied. Only few estimation methods can deal with imperfect data to a certain extent (e.g., Optimized Set Reduction, OSR®). Results from evaluating these methods in practical environments are rare. This article describes a case study on the application of OSR® at Toshiba Information Systems (Japan) Corporation. An important result of the case study is that estimation accuracy significantly varies with the data sets used and the way of preprocessing these data. The study supports current results in the area of quantitative cost estimation and clearly illustrates typical problems. Experiences, lessons learned, and recommendations with respect to data preprocessing and data-intensive cost estimation in general are presented.*

## 1. Introduction

Reliable software cost estimation is a crucial factor impacting project success. However, many software and system organizations still have significant problems in proposing realistic software costs, work within tight schedules, and finish their projects on schedule and within budget [26]. Considerable research has been directed at gaining a better understanding of the software development processes, and at building and evaluating cost estimation techniques, methods, and tools [9].

Recently, data-intensive estimation methods (that make intensive use of data to compute estimates) have been gaining more and more interest from both research and industry communities [1]. One reason is the increasing availability of data in industry that have been collected in a systematic way (e.g., motivated by companies' efforts to reach higher maturity levels). Organizations often want to gain more benefits from that measurement data. In addition, the cost-intensive involvement of experts in the cost estimation process could be reduced if data-intensive methods would reliably support the estimation process. Applying data-intensive estimation models also helps to get more insight on which factors (project characteristics) are cost-related. This could be used for initiating improvement programs that address those factors in future projects.

A major challenge in using data-intensive estimation methods is that the available data sets are typically not suitable for automated data analysis without initial analysis and preprocessing (because data are incomplete, partially invalid, or inconsistent) [18]. In practice, a common preprocessing strategy is to remove incomplete and inconsistent data items (e.g., the whole case is removed if one attribute value is missing). Alternatively, experts are involved to complete missing data. In consequence, significant parts of measurement data are either not considered at all or (in the best case) completed with subjective estimates. A few estimation methods exist that are able to deal with such imperfect data sets. Such methods usually do not make assumptions regarding data distribution and contain built-in mechanisms to handle missing data and data inconsistencies. However, evaluation results of such methods applied to recent data from industry (e.g., the comparison of methods described in [6]) are quite rare.

The objective of this article is to present empirically-based lessons learned and recommendations for dealing with imperfect industrial data sets for the purpose of cost estimation. The lessons learned and recommendations are derived from a case study that was conducted with Toshiba Information Systems (Japan) Corporation (TJ) in the context of a cooperation project between the Software Engineering Center of the Japanese Information-technology Promotion Agency (IPA-SEC) and the Fraunhofer Institute for Experimental Software Engineering (IESE). The estimation technique Optimized Set Reduction (OSR[®,1]) [1] was selected for the study, because it copes with numerous practical problems of industrial data sets. For example, it does not make any assumptions about the distribution of the underlying data, copes with missing data, and can operate on project characteristics on a nominal and continuous scale (like application type and effort, respectively). Moreover, the OSR® algorithm itself is completely automated. Those requirements were explicitly stated by TJ when selecting an appropriate algorithm.

The article is structured as follows: Section 2 introduces the relevant principles of the OSR® method. Section 3 presents the goals and the context of the industrial case study, its execution, analysis, and results, as well as a discussion of the validity. Section 4 describes lessons learned from the case study with respect to applying OSR® to an industrial data set. Section 5 discusses related work. Section 6 concludes with a list of recommendations for performing data-intensive cost estimation and illustrates future research directions.

## 2. The OSR® Method

The Optimized Set Reduction method (OSR®) is a pattern recognition method that analyzes trends in software engineering data sets based on machine-learning algorithms. An overview of the basics can be found in [8]. The idea is to select a subset of similar projects from a project data set as the basis for estimating a certain project attribute (like the overall effort of the project). The original set of project data is iteratively sub-divided into smaller sets, until a certain stop criterion is reached. Each project is described by a set of project characteristics, the so-called independent variables. Based on these independent variables, the sub-division is performed. The final sub-set is described by a Boolean expression, the so-called OSR® model, which is composed of independent variables that were identified as having a great influence on the variable to be estimated. The latter directly depends upon certain characteristics of the project data set and is therefore called dependent variable. The projects included in the set identified by the OSR® model are used to compute an estimate for the dependent variable. The algorithm may be used with different parameter settings (e.g., the function used to compute the final estimate, the function to assess the predictive power, and the stop criterion) that influence estimation accuracy. So, different combinations have to be evaluated in order to optimize estimation results for the data set of a specific organization. An overview of the overall approach is presented in Figure 1.

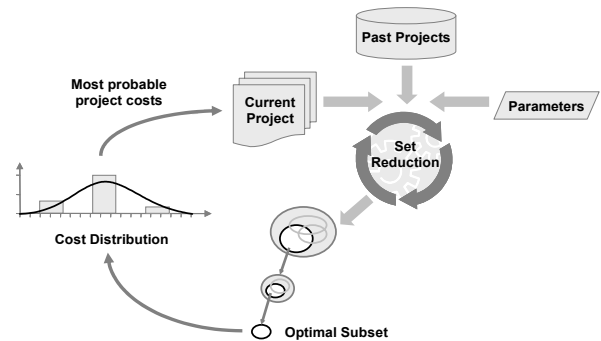

**Figure 1. Overview of the OSR® method.**

OSR® applications on real industrial data sets have shown that estimation accuracy is comparable to other data-intensive techniques (like Analogy-based) with one of the best standard deviations [9], [10]. But, in contrast to other techniques, OSR® was able to produce over 50% more predictions (for projects having missing data). The analyzed data sets were provided by the European Space Agency (ESA) and Laturi. The ESA data set was multi-organizational and included 160 projects (90 having complete data) from 4 companies from 1986 to 1998, with 17 characteristics per project. The Laturi data set included 206 software projects from 26 companies from 1986 to 1994, with 8 characteristics. OSR® was able to achieve an estimation accuracy of about 30% Mean Magnitude of Relative Error (including estimates for projects having incomplete data).

Benefits of OSR® include that it is able to work with missing data, provides means for uncertainty evaluation, is able to process continuous and discrete data, is automated, and produces well-interpretable outputs (OSR® models). These models contain several project characteristics and are not necessarily based mainly on a size measure (as is the case for COCOMO, for example). Recently, OSR® was applied to data sets of two Japanese companies (~80 and ~550 projects from different business units). The TJ case study applying OSR® and analyzing the outcomes will be described in more detail in the next section.

---
[1] OSR® is a registered trademark of the Fraunhofer Institute for Experimental Software Engineering.

## 3. The TJ Case Study

The case study was conducted by Fraunhofer IESE in collaboration with IPA-SEC and Toshiba Information Systems (Japan) Corporation (TJ). TJ provided a data base containing project data from different application domains within their organization. IPA-SEC and IESE analyzed the data set in terms of whether it is suited for data-intensive cost estimation. For this purpose, the OSR® algorithm was chosen to compute estimates for all projects and evaluate their quality using a cross-validation approach. The case study was divided into a pre-study and the application phase. The pre-study focused on evaluating the "technical" applicability of OSR® on industrial data provided; e.g., whether data quality allows applying the algorithm. The application phase evaluated the quality of OSR® estimates with respect to the estimation accuracy measured in terms of the Mean Magnitude of Relative Error (MMRE), Mean Squared Deviation (MSD), and Mean Absolute Deviation (MAD) as commonly used measures to evaluate the precision of estimation methods [12].

### 3.1. The OSR® Pre-Study

The goal of the pre-study was to define a list of transformation steps that were needed in order to apply OSR® and allow automated data analysis in general. For this purpose, TJ provided an initial data set of 78 projects. The OSR® method was applicable for the provided data sets and able to produce estimates with an MMRE of 37.01%. However, several issues were identified that required data preprocessing before the data set was suited for automatic processing. Besides some syntactical standard adaptation according to the OSR® input format, semantic transformation of several project variables was needed. This included unification of data values (actually string identifiers) representing the same category of a nominal-scale project characteristic. For instance, if a characteristic called "operating system" is represented by two different strings "Win2000" and "Windows 2000", we need a unique identifier so that OSR® does not assign them to different categories. This could also lead to separating one project characteristic into two. For instance, if the "version number" of an operating system should be separated from the "type", two new characteristics would replace the old "operating system" characteristic. A corresponding data preprocessing step would map different strings representing the same thing to a unique string. For the TJ case study, a mapping list was defined containing the old and new string representations, and was discussed with TJ experts.

### 3.2. The OSR® Application Phase

The goal of the OSR® application phase was to use the results (i.e., the list of transformation steps) from the pre-study and to apply the algorithm to an updated data set trying to achieve the best estimation accuracy possible and coming up with a list of lessons learned that will have to be considered when doing data-intensive cost estimation in general and estimation with OSR® in particular. The following issues were especially addressed: (A) Improved data set preparation (syntactically and semantically). (B) Clustering of data. (C) Comparison of OSR® results with standard regression analysis.

TJ prepared an updated data set (in comparison to the data set used in the pre-study) in order to be able to apply the OSR® tool suite and create OSR® estimates. After that, the independent variables and the dependent variable (the one that will be estimated) were determined. The "normalized performance index"[2] (which is a measure for determining a project's productivity) was chosen as the dependent variable, because it was seen as the main project planning criterion by the TJ people. Then, the OSR® tool suite was invoked using a direct cross-validation strategy; this means that certain subsets of a project data set were used as test set. For each project in the test set, an effort estimate was calculated using the rest of the projects (not included in the test set). After that, the estimated value was compared with the actual one provided in the project data set. This approach is used for computing the OSR® estimation accuracy by computing the Mean Magnitude of Relative Error (MMRE), the Mean Squared Deviation (MSD), and the Mean Absolute Deviation (MAD). OSR® can be used with different parameters and options (see [6]). In order to find the most suitable ones, the prediction accuracy was computed for differ-ent combinations of parameters and options. In order to get an impression of the quality of the OSR® results, we applied a linear regression approach (LRA) to the same test sets based on the "adjusted function points count" [5]. Size was chosen for the regression because it is seen as one of the most popular cost drivers in data-intensive cost estimation. LRA was chosen because of its popularity and simplicity.

---

[2] The TJ performance index was measured as function points per person hours. For privacy reasons, the performance index was normalized. This was done by dividing it by the average performance index over all projects analyzed.

### 3.3. Performing OSR® Analyses

When analyzing a data set with OSR®, the following steps have to be performed: Step 1 (data preparation): In this step, data are pre-processed, so that an OSR® analysis can be conducted. For instance, special characters are replaced, unique column identifiers for project characteristics are introduced, categories for each project characteristic on a nominal scale are analyzed, and unique categories are introduced and mapped to the original categories, if necessary.

Step 2 (data selection): In this step, the projects that will be included in the OSR® analysis as well as independent and dependent variables are determined. For instance, outlier projects (in terms of functional size and productivity) are excluded from the analysis. Moreover, the projects are randomly assigned to test-sets for cross-validation. After that, some basic statistics are computed, characterizing the data set, such as number of independent variables, number of projects, and ratio of missing data.

Step 3 (OSR® analysis): This step determines the different parameter combinations that are used for the OSR® analysis. After that, the OSR® analysis is conducted accordingly. As mentioned before, OSR® can be invoked with different parameters and options that lead to (slightly) different estimation results. The dependent variable for the TJ case study is on a continuous scale. This implies that several OSR® parameters and options are already fixed. This includes the Classification and Regression Trees (CART) [6] algorithm for discretizing continuous variables and Bootstrap [14] for computing the difference between distributions. For some parameters, commonly used values were chosen (which are used to fine-tune the algorithm). The significance level was set to 5% and the number of Bootstrap draws was set to 1000. For the prediction function, the objective function, the set size, and the predicate size, 36 parameter combinations were applied for each data subset analyzed (see Table 1). Four different set sizes were evaluated. For the TJ data set, we set the maximal set size that is evaluated to 20, because it seemed not to be reasonable to include more than a quarter of all projects in one set. With respect to the predicate size, we evaluated three settings.

The pre-study had shown that the OSR® models created did not change significantly if more than 4 predicates were added per iteration of the OSR® algorithm.

Step 4 (evaluate results): In this step, the results of the OSR® analysis are evaluated. For each parameter combination, the estimation accuracy (MMRE, MSD, and MAD) is computed. The best parameter combination is identified and compared to linear regression results.

**Table 1. OSR® parameter combinations used.**

| Prediction Function | Mean, Median (with MAD only) |
|---|---|
| Objective Function | MMRE, MSD, MAD |
| Minimal Set Size | 5, 10, 15, 20 |
| Max. Predicate Size | 2, 3, 4 |

### 3.4. Project Data Set

The TJ data set consists of 78 projects and 82 characteristics per project (excluding the project identifier). As mentioned above, the "normalized performance index" was determined as the dependent variable (which will be estimated by the OSR® algorithm). Furthermore, we reduced the project characteristics that will be considered in the prediction algorithm to 30 selected independent variables. The variables were selected based on the following criteria: (a) Ratio of missing data across all projects. If 90% or more of the project data were missing, the characteristic was not selected. (b) Redundancy of characteristics (e.g., "actual effort", "function point count", and "performance index"). (c) Explanatory power regarding the dependent variable. For instance, if a certain project characteristic has the same values across all projects, it does not contribute at all to explaining the variances of the project performance index and, in consequence, can be excluded from the analysis. In addition, 14 independent variables were revised and adapted in order to reduce the number of categories and to get unique categories for all variables. The final data set had a missing data ratio of 7.6%.

For the TJ case study, five different subsets were analyzed. Data set "A" removes obvious outliers with respect to functional size and projects that had more than 60% missing characteristics. As presented in Figure 2, four projects were marked as outliers and/or extreme values. Those projects were removed from the initial data set in order to get a cleaned starting point for the first OSR® computations. All other data sets analyzed were based on data set "A".

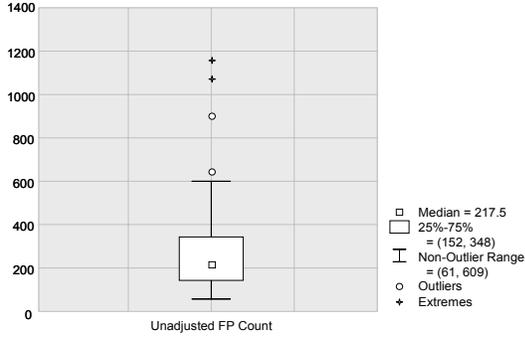

**Figure 2. Box plot of the unadjusted function point count.**

Data sets "B" and "C" try to reduce the range of the normalized performance index in order to see whether this would affect estimation accuracy. As shown in Figure 3, no outliers or extreme values could be detected by the analysis, but the overall range of the normalized performance index is quite large (0.2868 to 2.0581). However, 50% of all projects stay within a fairly small interval (0.6743 to 1.253). Data sets "D1" and "D2" were obtained by clustering the remaining data (data set "C") into new and enhancement projects, in order to see whether the development type has an effect on estimation accuracy. Table 2 presents an overview of the data sets, including the number of projects (#P), the number of characteristics (#C), and the ratio of missing data (MD).

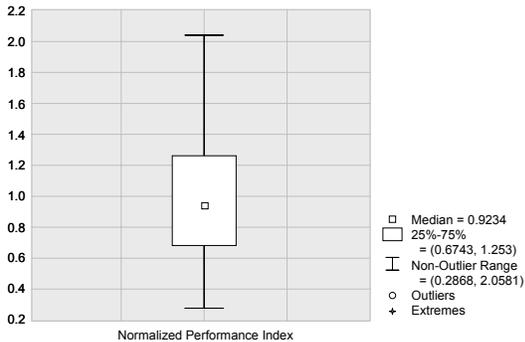

**Figure 3. Box plot of the normalized performance index.**

### 3.5. Analysis Results

For all analyzed data sets, the OSR® estimation accuracy was compared to a simple linear regression approach (LRA). We applied LRA using exactly the same projects for prediction as OSR®. The LRA estimates were computed based on the adjusted function point value for each project. As already mentioned, OSR® can be used with different parameters and options. Depending on the chosen options, the estimation accuracy can be quite different. This has something to do with the characteristics of the data set. Table 3 shows the estimation accuracy achieved with OSR® and linear regression. OSR® parameters that produced the best results are listed by referring to the prediction function, the objective function, the minimal set size, and the maximal predicate size. The estimation accuracy was determined using MMRE, MSD, and MAD measures over all estimated projects of the corresponding data subset.

**Table 2. Data set characteristics.**

|    | Data Set Description | #P | #C | MD |
|----|---|---|---|---|
| A  | No functional outliers | 72 | 30 | 5.56% |
| B  | Reduced productivity extreme values | 61 | 30 | 5.02% |
| C  | Even more reduced productivity extreme values | 58 | 30 | 5.23% |
| D1 | New development projects only | 36 | 30 | 5.64% |
| D2 | Enhancement projects only | 22 | 30 | 4.55% |

**Table 3. Case study results.**

|    | OSR® Parameters | | | | LRA | | OSR® |
|----|---|---|---|---|---|---|---|
| *Mean Magnitude of Relative Error* | | | | | | | |
| A  | Mean | MSD | 10 | 3 | 43.12% | > | 37.35% |
| B  | Mean | MSD | 10 | 3 | 30.79% | > | 27.19% |
| C  | Mean | MSD | 10 | 2 | 30.50% | > | 24.73% |
| D1 | Median | MAD | 5 | 2 | 37.04% | > | 21.75% |
| D2 | Median | MAD | 10 | 2 | 31.98% | > | 31.15% |
| *Mean Squared Deviation* | | | | | | | |
| A  | Mean | MMRE | 10 | 2 | 0.209 | > | 0.191 |
| B  | Mean | MMRE | 20 | 3 | 0.120 | < | 0.122 |
| C  | Mean | MMRE | 10 | 2 | 0.119 | > | 0.115 |
| D1 | Median | MAD | 5 | 2 | 0.181 | > | 0.091 |
| D2 | Median | MAD | 10 | 2 | 0.156 | > | 0.147 |
| *Mean Absolute Deviation* | | | | | | | |
| A  | Mean | MMRE | 10 | 2 | 0.367 | > | 0.358 |
| B  | Mean | MMRE | 15 | 2 | 0.284 | < | 0.287 |
| C  | Mean | MMRE | 10 | 2 | 0.284 | > | 0.272 |
| D1 | Median | MAD | 5 | 2 | 0.331 | > | 0.231 |
| D2 | Median | MAD | 10 | 2 | 0.315 | > | 0.311 |

In general, OSR® produces much better results than regression for MMRE in any case. However, with respect to the MSD and the MAD, OSR® produced less accurate results for data subset "B". The more outliers and extreme values are removed, the higher the estimation accuracy (from data set "A" to "C"). For different clusters of projects (new and enhancement in data sets "D1" and "D2"), quite different estimation accuracies were achieved. For heterogeneous data sets (depending on the variation, e.g., in productivity), OSR® produces much better estimates. The enhancement projects seem to be homogeneous with respect to the remaining characteristics considered in the analysis and thus, regression analysis produced nearly the same results as OSR®. For new development projects, the difference in estimation accuracy was huge. This data set seems to profit the most from OSR®.

Thus, OSR® accuracy (MMRE) improved significantly from data sets "A" to "C" and seems to mainly depend on the project type (see data sets "D1" and "D2"). The chosen OSR® parameters have something to do with the characteristics of the data set. When taking into account all data (including new and enhancement projects), nearly the same parameter combination was used (for runs "A", "B", and "C"). For new and enhancement projects, different parameter combinations produced the best results. For a certain type of data set, individual cross-validation has to be performed in order to identify the best combination.

### 3.6. Threats to Validity

The Magnitude of Relative Error (MRE) is one of the most common measures used to evaluate the precision of an estimation method. However, as stated by several researchers, it has several significant limitations when applied to compare estimation methods [15], [22]. In the case study presented here, it is mainly used to illustrate the improvement of OSR® when using different data sets. Its expressiveness with respect to comparing LRA and OSR® is limited.

Removing outlier projects and projects having extreme values (with respect to productivity) makes the data set more homogeneous and naturally produces better estimates. In our case study, linear regression improved, as did OSR®. For other data-driven estimation techniques and other environments, results may be different. However, the identified results give a first indication, even for problems that have to be considered using other data-driven estimation techniques.

### 4. Lessons Learned

The following section describes basic lessons learned from the case study regarding what has to be considered when doing data-intensive estimation with OSR® using industrial data. Some of them may seem obvious and may hold for many data-driven estimation methods. However, the following lessons learned illustrate those findings from a practical viewpoint.

(LL1) Size-based Cost Estimation: Considering other project characteristics than size helps to improve estimates and produce promising results, especially when dealing with inhomogeneous data. In our case study, the difference in estimation accuracy was quite large between OSR® and linear regression, which is based solely on size.

(LL2) Data Collection Process: The quality of the data collection process is essential if data-intensive estimation techniques are to be applied. Consistently specified scales of non-continuous project characteristics, for instance, facilitate reliable automatic analyses. Mapping and preprocessing project characteristics that are used for estimation is required for automated data analysis. If a column contains nominal values, the possible categories are checked and their names corrected. If necessary, the complete column has to be recorded in order to reduce the number of different categories for a certain project characteristic. In this case, the old categories have to be mapped to new ones. For instance, if there are 100 projects and 50 different categories for a project characteristic, about two projects will share one category (if the categories are equally distributed). If the minimum set size is 10, this characteristic would, for example, never be used in an OSR® analysis. Therefore, it is necessary to find more general categories by removing some details from the category definition (e.g., "Windows" instead of "Win2000"). In future applications, this mapping process should be reviewed in more detail by organization-internal experts to check whether the mapping is okay or whether a different mapping is more suitable.

(LL3) OSR® Parameter Selection: Finding the right OSR® parameters is a difficult task, but crucial for achieving good estimates. Currently, this is done in an exploratory approach, where the parameter combination leading to the best estimates is selected. This process is time-consuming, does not consider all parameter combinations and, therefore, does not guarantee optimal results. In the future, more guidance should be provided on how to determine the right parameter settings.

(LL4) Maintaining Data Sets: When using OSR® for predicting actual projects, it is important to update the estimation data base used regularly, so that estimates do not (solely) depend on projects that are older than, for instance, five years (depending on the organization). It is also important to detect outliers in the data set in order to improve estimation accuracy. This can be done using conventional mechanisms (like having a look at distributions and identifying outliers and extreme values). In the future, it could also be possible to use OSR® itself to detect this kind of outliers. As part of a different case study, we used an approach like this to see whether the data set will contain projects that are more similar and less outliers. This approach could be evaluated further in the future.

(LL5) Missing Data: As other studies have shown [11], the ratio of missing data has to be reduced in order to get good results. This is also supported when comparing the estimation accuracy of our pre-study and the main OSR® analyses.

(LL6) Reducing Scope: Clustering data with respect to selected project characteristics may improve the accuracy of estimates. For different clusters of projects ("D1" and "D2"), quite different estimation accuracies were achieved in our case study.

(LL7) Acceptance of Estimation Method: Cost estimation methods shall *support* a project planner in coming up with a reliable estimate and not replace her/him. Therefore, it is important that people applying a cost estimation method can trust the results and interpret them accordingly. In the TJ case, a number of practical problems had to be addressed before the method could be accepted. This included data storage issues, automated tool support, localization issues, proper identification of outliers, and the computation of confidence intervals for estimates.

## 5. Related Work

Numerous types of estimation methods have been developed over the last decade [10]. They basically differ with respect to the type of data they require and the form of the estimation model they do provide. With respect to input data, we differentiate between three major groups: data-intensive, expert-based, and hybrid methods (combining available data and expert knowledge in order to come up with estimates). The current trend among software organizations to increase the maturity of their software processes pushes software industry toward quantitative collection of measurement data. In parallel, data-intensive cost estimation methods are gaining more and more interest. The relatively high accuracy and the low application cost of data-intensive methods presented in related literature (as compared to traditionally acknowledged expert-based methods) are tempting for commercial software organizations that plan to collect or already do collect quantitative project data. Results presented in related literature do not provide a clear answer to the basic question of which method should be applied in a certain application context. The first impression when reviewing the numerous empirical studies published so far [19] is that the only reasonable criterion for evaluating an estimation method is its estimation accuracy. The second impression is that this criterion is probably not very helpful when selecting the most appropriate method, because the reader has to cope with contradicting outcomes of empirical investigations. For instance, [1] and [28] present contradicting results, although they evaluate the same data set.

One of the most significant issues neglected in many empirical studies is the applicability of an estimation method in a certain context. In case of data-intensive methods, this includes the required quantity and quality of project data. In practice, even if measurement processes are defined and in place, if data are not collected according to a previously explicitly defined measurement goal, the probability is quite high that data are inconsistent and/or incomplete. Experts, on the other hand, vary largely when assessing non-continuous data. This leads to highly incomplete data sets with numerous irrelevant factors and data outliers. Efficient estimation methods pretending to support software practitioners should cope with all those problems. Moreover, they should provide appropriate decision support facilities (e.g., estimate uncertainty evaluation).

Among the data-intensive methods, some require past project data in order to build customized models (*define-your-own-model* approaches), others provide an already defined model, where factors and their relationships are fixed (*fixed-model* approaches). The popular COCOMO models [1], [3] or SLIM [20] are examples of fixed models. (COCOMO also allows calibrating the initial model with one's own project data.) The major advantage of fixed-model approaches is that they, theoretically, do not require any data from already completed projects. Those methods are especially attractive to organizations that have not started collecting quantitative data yet. Yet, in practice, fixed models are developed for a specific context and are, by definition, only suited for estimating the types of projects for which the fixed model was built. The applicability of such models for different contexts is usually quite limited. In order to improve their performance, organization-specific project data are required anyway for calibrating the generic model in a specific application context. Moreover, a fixed model requires a specific set of project characteristics (so-called cost factors) to be measured. This often includes a fixed size measure (e.g., only source lines of code). Fixed models may include factors that are irrelevant in a certain context, while excluding others having a significant im-pact on project cost. In contrast, define-your-own-model approaches use organization-specific project data in order to build a model that fits to a certain organizational context. These methods require neither any specific set of factors to be measured nor a certain size measure. Parametric statistical methods, such as those based on regression [7] or analysis of variance (ANOVA) [16], make assumptions regarding the distribution of the underlying data, which can almost never be fulfilled in practice. Regression models additionally require defining an a priori functional form of the cost function, which requires large data sets, do not perform well with discontinuous variables, and are very susceptible to the effect of outliers [23].

Non-parametric methods originating from the machine learning domain such as artificial neural networks (ANN) [4], Classification and Regression Trees (CART) [6], or Optimized Set Reduction (OSR®) [8], have recently gained much interest among researchers, since they make practically no assumptions about the data and can deal with mixed continuous/non-continuous and messy data [25]. They are, however, quite sensitive to their parameter configuration [24] and there is usually little universal guidance regarding how to set those parameters. Thus, finding appropriate parameter values requires some preliminary experimentation.

Analogy-based methods such as Case-based reasoning (CBR) [24] implement an estimation mechanism similar to that used by human estimators and are therefore more intuitive. They take already finished projects that are similar to the one to be estimated in order to come up with an estimate. However, similarity measures are intolerant of noise and of irrelevant cost factors. This might be dealt with by applying additional factor selection techniques [1].

Finally, there are several data-related problems that affect nearly all data-intensive cost estimation methods. Missing data, for instance, is a very common weakness of industrial data sets that has a significant impact on the applicability of the method. The OSR® application as described in this paper explicitly addresses data preprocessing steps in order to deal with missing and messy data. Handling missing data is not new in the data analysis domain [17], [21]. However, only few approaches have already been applied in the software engineering domain in general, and to cost estimation in particular (e.g., [11]). Another common problem is the difficulty that a project manager faces when having to specify values for unknown qualitative cost factors. Handling those data in terms of triangular distribution [27] or (most recently) as fuzzy numbers [13] is proposed as a solution. If only the final estimate is presented, false conclusions may be drawn in terms of confidence in estimation accuracy. While confidence intervals can be developed, this is rarely done, and given the small data sets available (with skewed distributions), the intervals are often questionable.

In summary, software decision makers face numerous practical problems when applying data-intensive cost estimation methods. Software estimators need support to select and apply data preparation and cost estimation methods. This calls for practical guidelines and reliable field studies regarding the application of such methods in industrial environments, on up-to-date project data.

## 6. Conclusion and Future Work

Data-intensive methods have numerous advantages. Yet, in order to fully benefit from the application of such methods, the data have to have an appropriate quality and quantity. In our study we have applied a data-intensive method, OSR®, which was designed to cope with most of the common problems of industrial data, such as missing data or mixed continuous and discontinuous data. We found that there are still a number of issues to be solved that might affect the quality of predictions when applying data-intensive estimation methods. These issues include:

- Data quality has to be addressed. Before employing automated estimation of project cost, the quality of input measurement data has to be carefully analyzed. Missing data must be resolved and outliers must be detected and removed before acceptable estimation results can be obtained. In addition, consistently defined unified measurement scales are needed for non-continuous factors and the right project characteristics have to be measured, i.e., only those having a significant influence on observed project characteristics such as productivity or cost.

- Data quantity has to be addressed. Data-intensive estimates can support a planner in coming up with a prediction. However, it depends on the number of similar projects how reliable such estimates are. There is also a technical issue related to data quantity. Advanced data-intensive estimation techniques use computationally intensive algorithms (in particular, machine learning approaches). The more data are analyzed, the longer it takes to come up with an estimate.

The results of the case study support current knowledge in the area of quantitative cost estimation and lead to the following recommendations:

- An organization may profit by collecting more than "just" size when doing data-driven estimation.

However, it is therefore important to know which factors are important for an organization or a certain part of an organization. Goal-oriented measurement [2] can support this process by systematically taking into account influencing factors and coming up with reliable measures.

- As the study indicates, it can help to improve estimation accuracy if estimates are computed for a smaller scope of projects with more homogeneous characteristics (e.g., through data clustering).

- Hybrid data- and expert-based identification of the most significant influencing factors and relationships between them can help to focus the data collection process, improve prediction accuracy, and reduce costs. It would also be possible to develop a kind of causal model (as done for the CoBRA® method [27]) that explicitly describes the interaction between project cost and influencing factors by making use of experts.

- A systematic, restrictive data collection (and measurement) process is needed for doing data-intensive cost estimation. Otherwise, intensive rework and preprocessing is needed before being able to do cost estimation. Moreover, the estimation model needs to be maintained over time. So, processes have to be in place on how to update the estimation base with new projects and how to guarantee a certain quality of the estimation model (model validation).

- Data-intensive estimation methods (like OSR®) usually make use of quite complex algorithms and therefore need tool support. This support should, however, not be restricted to the algorithm itself. When introducing such a method to an organization, it is also important to support the organizational processes around the pure application of the algorithm. This includes aspects like maintaining the estimation base, detecting outliers, and controlling the quality of incoming data syntactically as well as semantically.

Project planning is a human-based process and estimation methods should support project planners and decision makers and not replace them. In that sense, any additional information provided by a method such as estimation uncertainty or information about the estimation model (included factors and dependencies) can help to understand the estimates obtained and the related software processes. This can be a fundamental factor for the practical usefulness of an estimation method.

Future work will focus on the following aspects:

- We compared OSR® exclusively against linear regression, a quite simple estimation approach. For future studies, it would be interesting to compare results and issues with other data-intensive estimation methods in an industrial context.

- The selection of project characteristics is important when dealing with many independent variables. OSR® is a computationally intensive algorithm. The maximal number of predicates used in an iteration of the OSR® algorithm has a strong influence on computation time. For the OSR® application phase, the selection of characteristics was basically done by external experts (IPA-SEC and Fraunhofer IESE). In the future, this process should mainly be driven by internal experts and could also be supported by quantitative data analysis, so that the benefits of data-driven cost estimation can be maximized.

- Finally, guidelines should be derived on how to define measurement processes for data-intensive cost estimation, how to select the right estimation method for different organizations (best practices), how to prepare data, how to apply the estimation methods, and finally, how to maintain the estimation base and the estimation model.

## Acknowledgement

We would like to thank Toshiba Information Systems (Japan) Corporation, where we conducted the study, as well as all involved experts and local organizers, who greatly contributed to the successful performance of the project. We would also like to thank the Japanese Information-technology Promotion Agency (IPA) for their personnel and financial support in conducting the case study. Finally, we would like to thank Sonnhild Namingha and Michael Klaes from Fraunhofer IESE for reviewing a first version of the article.